\documentclass[twocolumn,preprintnumbers,amsmath,aps]{revtex4}
\usepackage{graphicx}
\usepackage{dcolumn}
\usepackage{bm}
\usepackage{subfigure}
\usepackage{color}
\begin{document}
%%%%%%%%%%%%%%%%%%%%%%%%%%%%
\newcommand{\kvec}{\mbox{{\scriptsize {\bf k}}}}
%%%%%%%%%%%%%%%%%%%%%%%%%%%%
\def\eq#1{(\ref{#1})}
\def\fig#1{figure\hspace{1mm}\ref{#1}}
\def\tab#1{table\hspace{1mm}\ref{#1}} 
%%%%%%%%%%%%%%%%%%%%%%%%%%%%
\title{On the high-pressure superconducting phase in platinum hydride}
\author{D. Szcz{\c{e}}{\'s}niak$^1$}\email{dszczesniak@qf.org.qa}
\author{T. P. Zem{\l}a$^2$}
%%%%%%%%%%%%
\affiliation{$^1$Qatar Environment and Energy Research Institute, Qatar Foundation, PO Box 5825, Doha, Qatar}
\affiliation{$^2$Institute of Physics, Jan D{\l}ugosz University in Cz{\c{e}}stochowa, Ave. Armii Krajowej 13/15, 42-200 Cz{\c{e}}stochowa, Poland}
%%%%%%%%%%%%
%%%%%%%%%%%%
\date{\today} 
\begin{abstract}
%%%%%%%%%%%%%%%%%%%%%%%%%%%%%%%%%%%%%%%%%%%%%%%%%%%

Motivated by the ambiguous experimental data for the superconducting phase in silane (SiH$_{4}$), which may originate from the platinum hydride (PtH), we provide a theoretical study of the superconducting state in the latter alloy. The quantitative estimates of the thermodynamics of PtH at 100 GPa are given for a wide range of the Coulomb pseudopotential values ($\mu^{*}$) within the Eliashberg formalism. The obtained critical temperature value ($T_C\in\left<12.94, 20.01\right>$ for $\mu^{*}\in\left<0.05,0.15\right>$) agrees well with the experimental $T_{C}$ for SiH$_{4}$ which may be ascribed to PtH. Moreover, the calculated characteristic thermodynamic ratios exceed the predictions of the Bardeen-Cooper-Schrieffer theory, implying occurrence of the strong-coupling and retardation effects in PtH. We note that our results can be of high relevance for the future studies on hydrides.

%%%%%%%%%%%%%%%%%%%%%%%%%%%%%%%%%%%%%%%%%%%%%%%%%%%
\end{abstract}
\maketitle

%%%%%%%%%%%%%%%%%%%%%%%%%%%%%%%%%%%%%%%%%%%%%%%%%%%
\section{Introduction}
%%%%%%%%%%%%%%%%%%%%%%%%%%%%%%%%%%%%%%%%%%%%%%%%%%%

Recent experimental results on the superconducting phase in H$_2$S hydride at high pressure \cite{drozdov}, which report record high critical temperature value ($T_{C} \sim 190$ K), may become an another breakthrough in the research on the room-temperature superconductivity. In addition to the exceptional thermodynamic properties of the H$_2$S alloy \cite{li}, \cite{durajski1}, \cite{errea1}, discussed material is considered to be a phonon-mediated superconductor. The latter fact is of great importance since electron-phonon superconductors can be described within well-established Bardeen-Cooper-Schrieffer theory (BCS) \cite{bardeen1}, \cite{bardeen2} and its derivatives such as the strong-coupling Eliashberg formalism \cite{eliashberg}.

In this spirit, hydride materials, whose concept stems from the milestone works by Ashcroft \cite{ashcroft1}, \cite{ashcroft2}, are currently one of the most promising candidates for the future high-temperature superconductors \cite{szczesniak1} in comparison with {\it e.g.} cuprates \cite{szczesniak2}, \cite{szczesniak3}. Among the family of hydrogen-based materials, the high-pressure superconducting phase has been experimentally observed only in the aforementioned H$_2$S alloy and likely in the platinum hydride material (PtH). Please note however that these results are not fully confirmed yet.

From the scientific point of view, the irrefutable testimonies for the existence of the superconducting phase in the PtH and H$_2$S alloys are equally important for pushing forward research in the discussed domain. However, in the case of PtH the already available results are not so clear and direct like in the case of the H$_2$S hydride.

In particular, in 2008 Eremets {\it et al.} conducted the high-pressure experiment for silane (SiH$_4$) and stated that the induction of the superconducting state in this material is possible \cite{eremets}. However, further investigations of the results obtained by Eremets {\it et al.} questioned to some extent the outcome of the discussed work. Degtyareva {\it et al.} claimed in their paper \cite{degtyareva} that, in the discussed experiment, silane can decompose and released hydrogen may react with the surrounding metals, in particular with platinum. This statement was motivated by the XRD data which matches PtH instead of the SiH$_4$ alloy, suggesting that the observed superconducting state may originate from the first mentioned material. Further theoretical studies confirmed that such scenario is highly possible \cite{errea2}, \cite{zhang}, \cite{zhou}, \cite{kim}. However, all mentioned theoretical papers presented approximate estimations of the critical temperature by using the McMillan formula \cite{mcmillan2}, and did not provide the discussion of other crucial thermodynamic properties. In this context, the discussed problem of the existence of the superconducting state in PtH is still an open and emerging question.

In order to address raised in the previous paragraph issues, we present complementary and comprehensive theoretical analysis of all of the most important thermodynamic properties of the PtH superconductor at the pressure of $p=100$ GPa. This pressure value is chosen in order to match the conditions at which the highest value of the critical temperature has been obtained in the experiment of Eremets {\it et al.} \cite{eremets}. In particular, our calculations are conducted within the Eliashberg theory due to the relatively high electron-phonon constant for the PtH hydride ($\lambda=0.84$ as predicted by Kim {\it et al.} \cite{kim}).

The present paper is organized as follows, in the first section we briefly elaborate on the technical aspects of the Eliashberg formalism and applied numerical procedures. Next we present our numerical results obtained on the imaginary axis as well as in the mixed representation of the Eliashberg equations. Our paper ends with some pertinent conclusions.

%%%%%%%%%%%%%%%%%%%%%%%%%%%%%%%%%%%%%%%%%%%%%%%%%%%
\section{Theoretical model}
%%%%%%%%%%%%%%%%%%%%%%%%%%%%%%%%%%%%%%%%%%%%%%%%%%%

The theoretical analysis presented in this paper based on the isotropic Eliashberg equations which are solved, at the imaginary axis and in the mixed representation, by using the iterative method given in \cite{durajski2} and \cite{szczesniak4}. The isotropic character of the Fermi surface is imposed by the Eliashberg spectral function ($\alpha^2F(\Omega)$), which models the electron-phonon interactions and was taken from \cite{kim}. We note, that in order to assure the consistency of our calculations with the experimental conditions in \cite{eremets}, the chosen $\alpha^2F(\Omega)$ function corresponds to the pressure value of 100 GPa and the {\it hcp} structure of the space group P$6_3$.

Due to the uncertain experimental data we consider three distinct values of the Coulomb pseudopotential ($\mu^{*}$) which models the depairing electron correlations and allows us to sample the superconducting phase of PtH. In particular, $\mu^{*}=\left( 0.05, 0.1, 0.15 \right)$ and match the general estimations suggested by Ashcroft for the hydride superconductors in \cite{ashcroft1}. Furthermore, the wide range of the $\mu^{*}$ values allows us to appropriately examine convergence of our results and the experimental data of Eremets {\it et al.} \cite{eremets}, as well as investigate the influence of the $\mu^{*}$ value on the superconducting properties of PtH.

The precision of our numerical analysis is controlled by assuming the sufficiently high limit for the Matsubara frequencies: $\omega_{m}\equiv\frac{\pi}{\beta}(2m-1)$, where $\beta\equiv 1/k_{B}T$ is the so-called inverse temperature, and $k_{B}$ states for the Boltzmann constant. Our calculations are conducted then for the 2201 Matsubara frequencies, allowing us to estimate quantitatively the superconducting properties of the PtH for temperatures $T\geq T_{0}\equiv 3$ K. Finally, the cut-off frequency is set to be $\omega_{c}=10\Omega_{\rm max}$, where $\Omega_{\rm max}$ is the maximum phonon frequency equals to $193.25$ meV.
 
%%%%%%%%%%%%%%%%%%%%%%%%%%%%%%%%%%%%%%%%%%%%%%%%%%%
\section{Numerical results and discussion}
%%%%%%%%%%%%%%%%%%%%%%%%%%%%%%%%%%%%%%%%%%%%%%%%%%%

\begin{figure}[ht]
\includegraphics[width=\columnwidth]{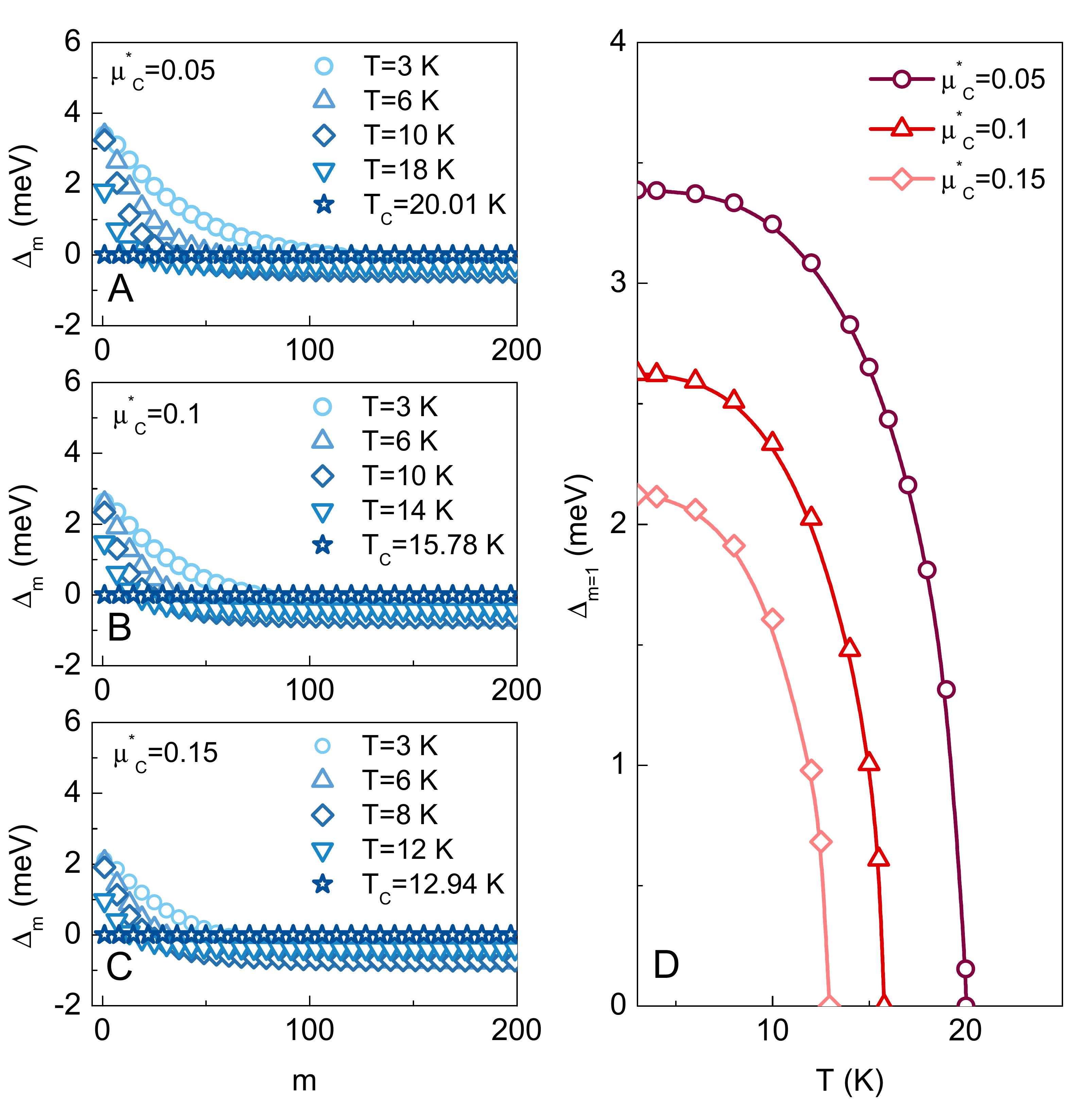}
\caption{(A)-(C) The order parameter ($\Delta_m$) on the imaginary axis as a function of $m$ for selected values of the temperature ($T$) and Coulomb pseudopotential ($\mu^*$). (D) The maximum value of the order parameter ($\Delta_{m=1}$) on the imaginary axis as a function of the temperature for the selected values of the Coulomb pseudopotential.}
\label{fig1}
\end{figure}

In the first step we analyze the temperature dependance of the order parameter ($\Delta_{m}$) by conducting our numerical calculations on the imaginary axis. The obtained results are presented, for all considered $\mu^{*}$ values and selected temperatures, in Figs. \ref{fig1} (A-C) as a function of $m$. The physically reasonable gradual decrease of the $\Delta_{m}$ value with increasing $\mu^{*}$ and $T$ can be observed. Moreover, together with increasing $m$ we notice saturation of the obtained solutions in all cases. It is a testimonial for the precision of our calculations and the quantitative character of the obtained results.

The value of the critical temperature is determined on the basis of the condition: $\Delta_{m=1}(T_{C})=0$, where $\Delta_{m=1}$ denotes the maximum value of the order parameter. Supplementing functional dependance of the $\Delta_{m=1}$ function on the temperature for the selected values of $\mu^{*}$ is depicted in Fig. \ref{fig1} (D). In particular, calculated $T_{C}$ amounts 20.01 K, 15.78 K, and 12.94 K for $\mu^{*}$ equals to 0.05, 0.1, 0.15, respectively.

We note that the experimental value of $T_{C}$ obtained by Eremets {\it et al.} \cite{eremets} is $\sim 17$ K for $p \approx 100$ GPa. In what follows, it matches our predictions falling in to the physically feasible region of $\mu^{*} \in (0.05, 0.1)$.

The results obtained on the imaginary axis and presented in Figs. \ref{fig1} (A-D) provide also information on other measurable observable, namely the effective electron mass ($m^{\star}_{e}$). Specifically, the value of the $m^{\star}_{e}$ can be determined by using the following relation:
\begin{equation}
\label{eq1}
m^{\star}_{e}\simeq Z_{m=1} m_{e},
\end{equation}
where $Z_{m=1}$ is the maximum value of the wave function renormalization factor and $m_{e}$ denotes the band electron mass.

\begin{figure}[h]
\includegraphics[width=\columnwidth]{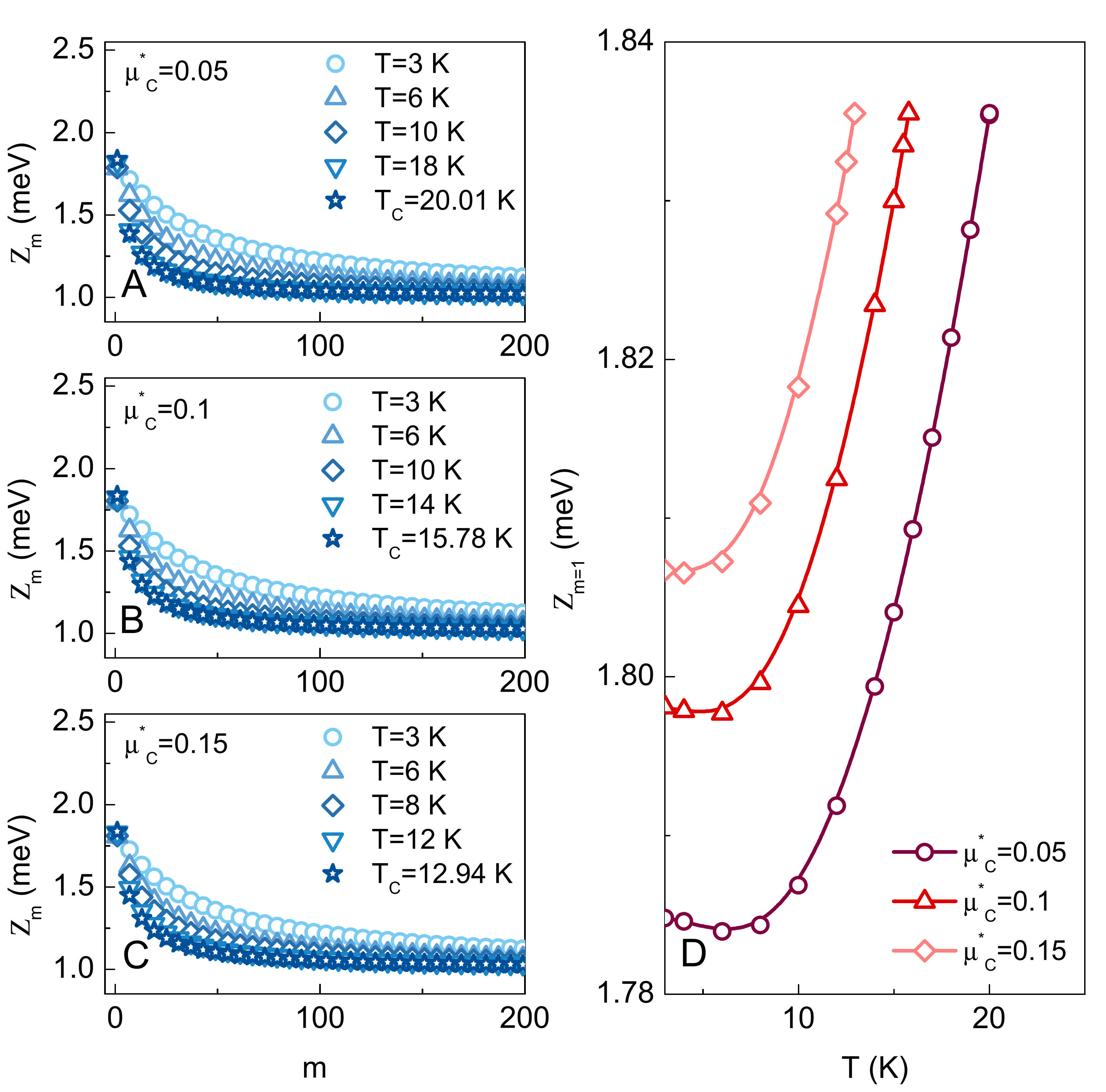}
\caption{(A)-(C) The wave function renormalization factor on the imaginary axis ($Z_{m}$) as a function of $m$ for the selected values of temperature ($T$) and Coulomb pseudopotential ($\mu^*$). (D) The maximum value of the wave function renormalization factor on the imaginary axis ($Z_{m=1}$) as a function of the temperature for the selected values of the Coulomb pseudopotential.}
\label{fig2}
\end{figure}

The wave function renormalization factor as well as its maximum value are presented in Figs. \ref{fig2} (A-D) as a functions of the temperature for the selected values of $\mu^{*}$. By using Eq. (\ref{eq1}) we estimate the $m^{\star}_{e}$ value for the PtH alloy to be 1.84 $m_{e}$, which is relatively high as for the phonon-mediated superconductors.

The determined dependance of the $Z_{m=1}$ function on the temperature allows us next to calculate the normalized free energy difference between the superconducting and normal state as:
\begin{eqnarray}
\label{eq2}
\frac{\Delta F}{\rho\left(0\right)}&=&-\frac{2\pi}{\beta}\sum_{n=1}^{M}
\left(\sqrt{\omega^{2}_{n}+\Delta^{2}_{n}}- \left|\omega_{n}\right|\right)\\ \nonumber
&\times&(Z^{S}_{n}-Z^{N}_{n}\frac{\left|\omega_{n}\right|}
{\sqrt{\omega^{2}_{n}+\Delta^{2}_{n}}}).
\end{eqnarray}  
In Eq. (\ref{eq2}), $\rho(0)$ states for the electron density of states at the Fermi level,  whereas $Z^{S}_{n}$ and $Z^{N}_{n}$ are the wave function renormalization factors for the superconducting ($S$) and normal ($N$) state, respectively.

\begin{figure}[ht]
\includegraphics[width=\columnwidth]{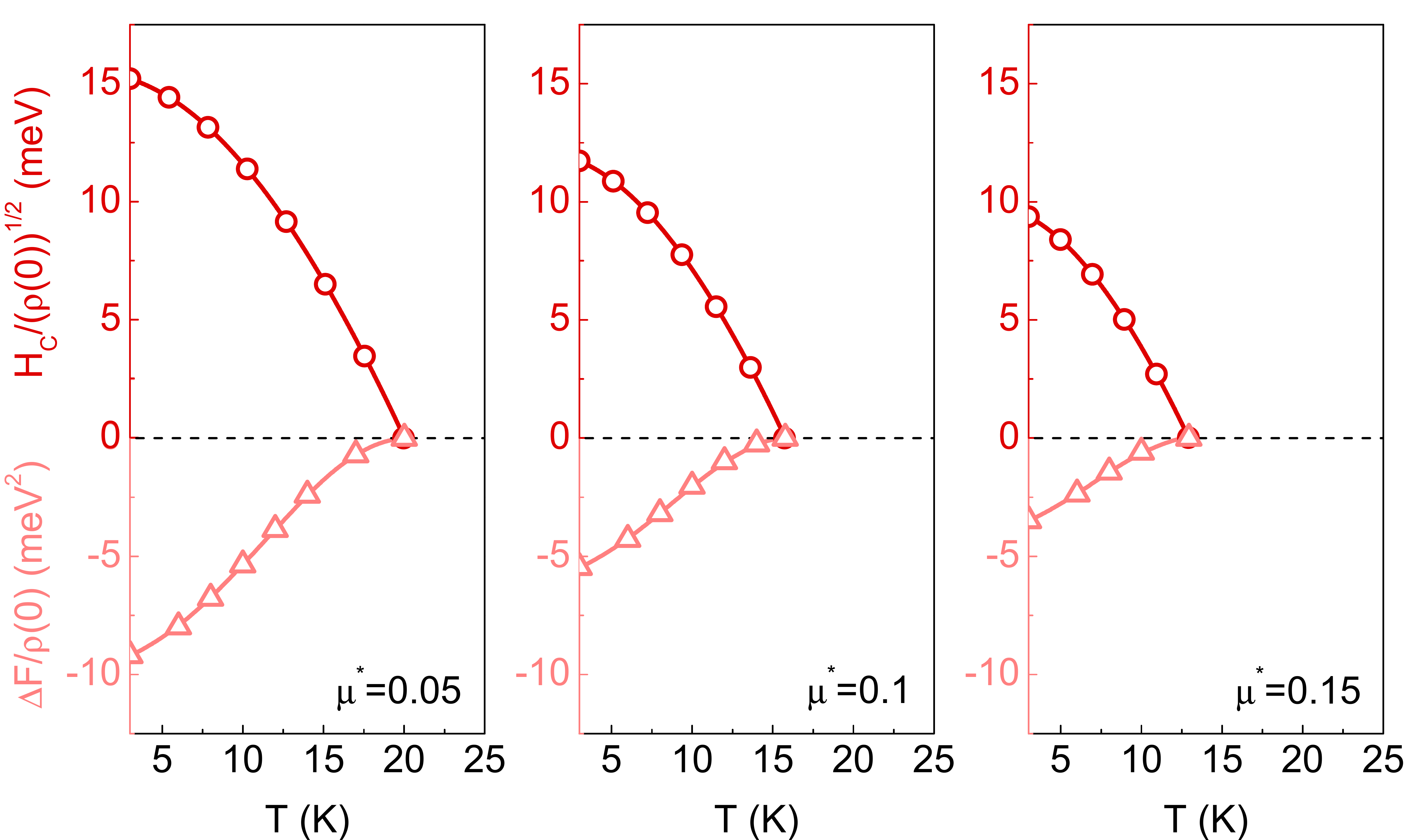}
\caption{(A-C) The normalized free energy difference ($\Delta F/\rho\left(0\right)$) (lower panel) and the thermodynamic critical field (${H_{C}}/{\sqrt{\rho\left(0\right)}}$) (upper panel) as a function of the temperature  ($T$) for the selected values of the Coulomb pseudopotential ($\mu^{*}$).}
\label{fig3}
\end{figure}

The $\Delta F/\rho\left(0\right)$ as a function of temperature for three considered $\mu^{*}$ values is presented in the lower panels of Figs. \ref{fig3} (A-C). We observe that in all cases $\Delta F/\rho\left(0\right)$ takes negative values indicating thermodynamic stability of the superconducting phase, where togheter with decreasing value of $\mu^{*}$ the stability increases.

Figs. \ref{fig3} (A-C) present also corresponding functional behavior of the normalized thermodynamic critical field ${H_{C}}/{\sqrt{\rho\left(0\right)}}$ on temperature, obtained on the basis of the following expression:
\begin{equation}
\label{eq3}
\frac{H_{C}}{\sqrt{\rho\left(0\right)}}=\sqrt{-8\pi\left[\Delta F/\rho\left(0\right)\right]}.
\end{equation}

Like in the case of the $\Delta F/\rho\left(0\right)$ function, the normalized thermodynamic critical field strongly depends on the assumed $\mu^{*}$ value, and decrease together with increasing $\mu^{*}$. The ${H_{C}}/{\sqrt{\rho\left(0\right)}}$ function and the specific heat for the normal state ($C^{N}$) given by:
\begin{equation}
\label{eq4}
\frac{C^{N}\left(T\right)}{ k_{B}\rho\left(0\right)}=\frac{\gamma}{\beta}, 
\end{equation}
where $\gamma\equiv\frac{2}{3}\pi^{2}\left(1+\lambda\right)$ is the Sommerfeld constant, can be next used to to estimate the first characteristic thermodynamic ratio, familiar in the BCS theory. In particular, the ratio of interest for the critical magnetic field is defined as:
\begin{equation}
\label{eq5}
R_{H}\equiv\frac{T_{C}C^{N}\left(T_{C}\right)}{H_{C}^{2}\left(0\right)},
\end{equation}
and takes the constant value of R$^{\rm BCS}_{\rm H}$=0.168 within the BCS theory \cite{bardeen1}, \cite{bardeen2}. However, for the PtH alloy at 100 GPa our calculations predict that R$_{\rm H} \in \left< 0.155,  0.167 \right>$ for $\mu^{*} \in\left< 0.05, 0.15 \right>$, showing deviation from the BCS predictions. 

\begin{figure}[h]
\includegraphics[width=\columnwidth]{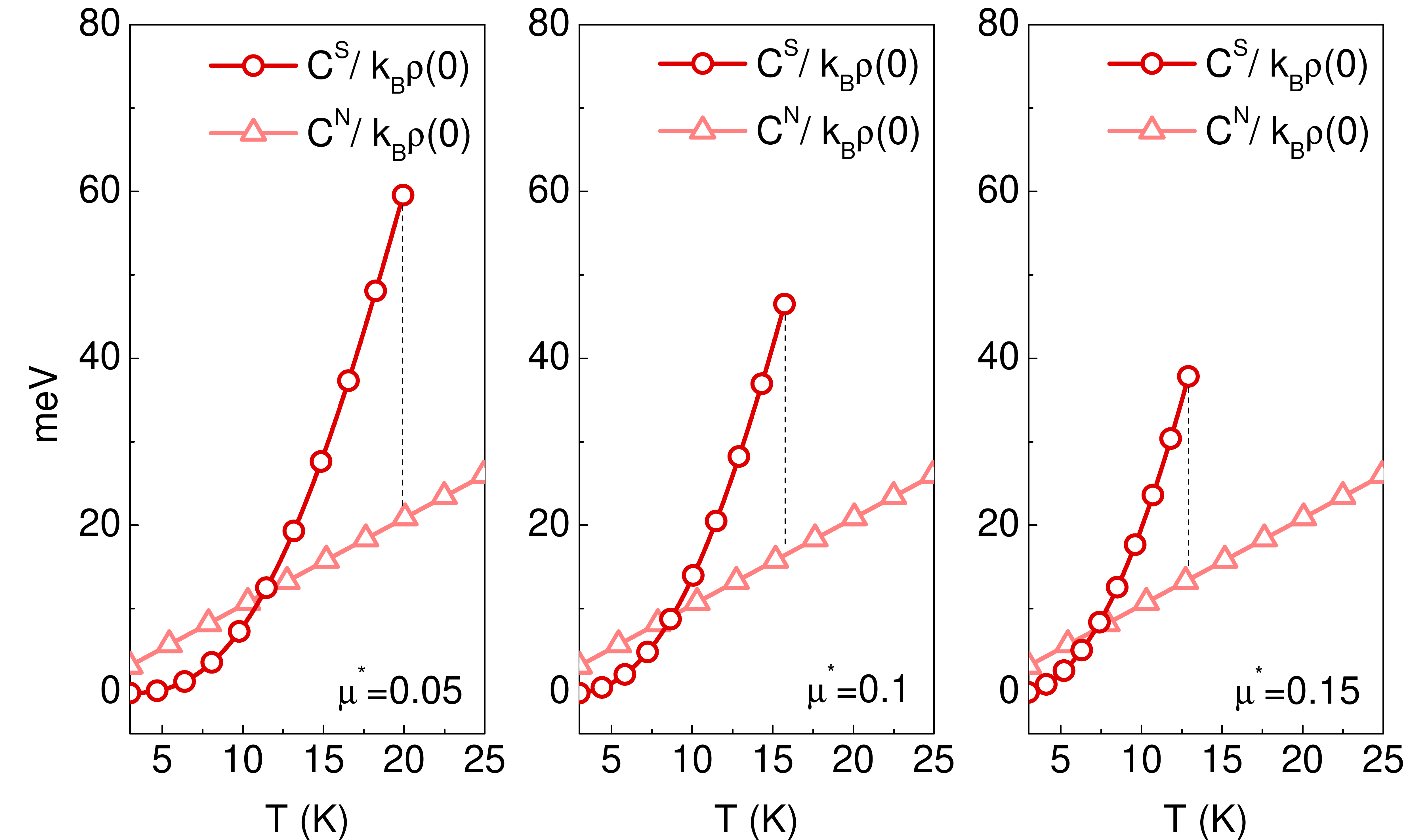}
\caption{The normalized specific heat of the superconducting ($C^{S}/{k_{B} \rho(0)}$) and normal ($C^{N}/{k_{B} \rho(0)}$) state as a function of the temperature ($T$) for the selected values of Coulomb pseudpotential ($\mu^*$).}
\label{fig4}
\end{figure}

Relations given in Eqs. (\ref{eq2}) and (\ref{eq4}), can be next used to address the normalized difference between the specific heat of the superconducting (C$^{S}$) and normal (C$^{N}$) state, as given by:  
\begin{equation}
\label{r12}
\frac{\Delta C\left(T\right)}{k_{B}c\rho\left(0\right)}=-\frac{1}{\beta}\frac{d^{2}\left[\Delta F/\rho\left(0\right)\right]}{d\left(k_{B}T\right)^{2}},
\end{equation}
where $\Delta C=C^{S}-C^{N}$. In what follows, the specific heat of the superconducting state can be determined. In Figs. \ref{fig4} (A-C) we present obtained functional behavior of the $C^{S}$ on the temperature for the three considered $\mu^{*}$ values. For comparison and systematical purposes, the $C^{N}$ function is also presented. It can be observed, that $C^{S}$ function rises faster then $C^{N}$ as temperature grows, presenting characteristic {\it jump} at $T_{C}$. Moreover, the normalized specific heat of the superconducting state strongly depends on the $\mu^{*}$ value, decreasing when the $\mu^{*}$ grows. 

For the results given in Figs. \ref{fig4},  the second dimensionless characteristic thermodynamic ratio can be calculated, by using:
\begin{equation}
\label{r14}
R_{C}\equiv\frac{\Delta C\left(T_{C}\right)}{C^{N}\left(T_{C}\right)}.
\end{equation}
Similarly like in the case of the $R_{H}$ ratio, the obtained $R_{C}$ values exceed the BCS limit. In particular, $R_{C} \in \left< 1.88, 1.82 \right>$ for $\mu^{*} \in \left< 0.05, 0.15 \right>$, whereas $R^{BCS}_{C}=1.43$.

The last remaining characteristic ratio, which corresponds to the energy gap value at the Fermi level ($2\Delta(T)$, where $\Delta\left(T\right)$ is the the order parameter on the real axis), can be calculated by solving the Eliashberg equations in the mixed representation (please see \cite{durajski2} and \cite{szczesniak4} for more details). Such analysis is required in order to provide the quantitative values of the $\Delta\left(T\right)$ function, given by the following expression:

\begin{equation}
\label{eq3}
\Delta\left(T\right)={\rm Re}\left[\Delta\left(\omega=\Delta\left(T\right),T\right)\right].
\end{equation}

\begin{figure}[h]
\includegraphics[width=\columnwidth]{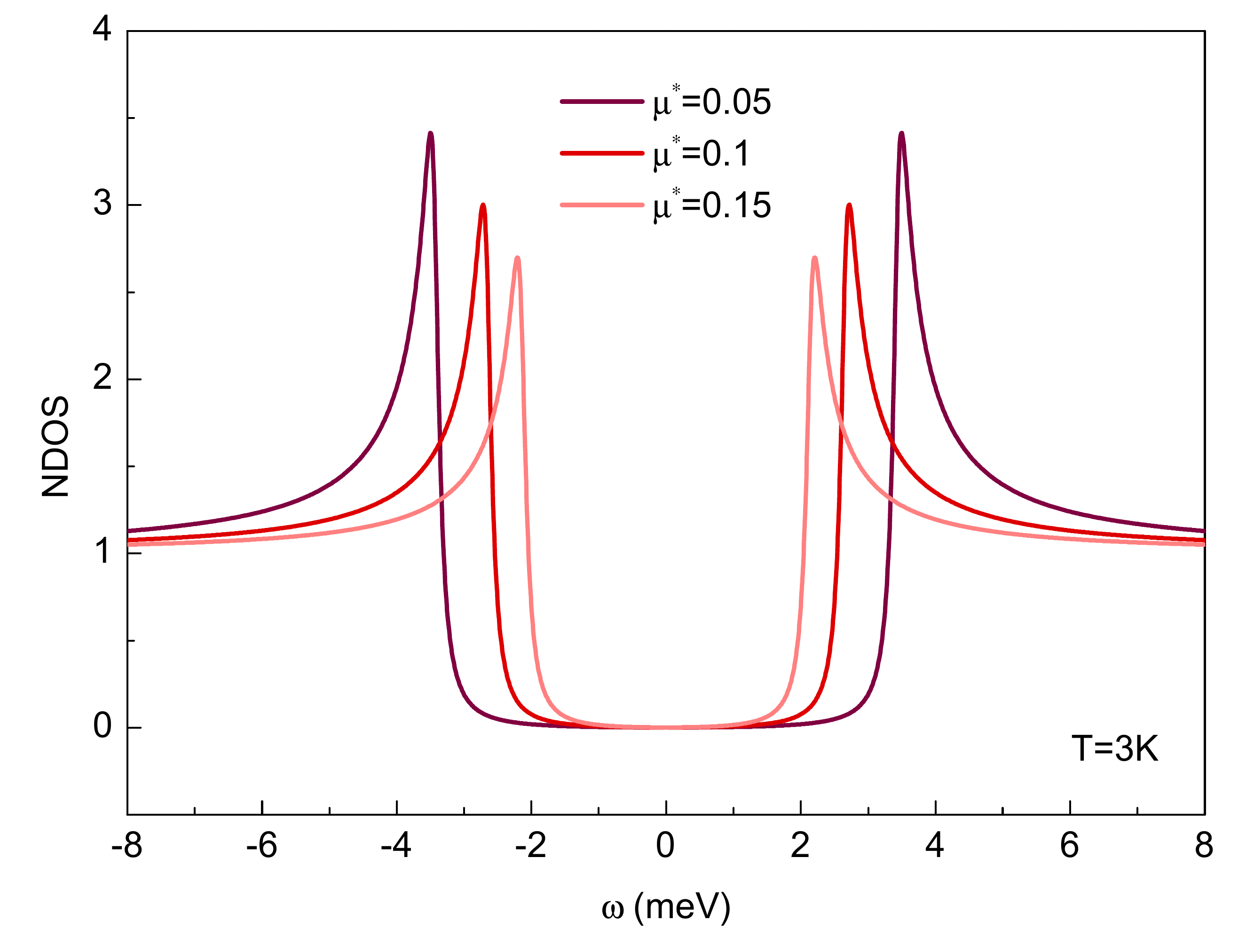}
\caption{The normalized total electronic density of states (NDOS) at $T_0=T=3K$ for the selected values of Coulomb pseudopotential ($\mu^*$).}
\label{fig5}
\end{figure}

In Fig. \ref{fig5}, we summarize our calculations on the real axis by presenting the results of the total normalized density of states (NDOS$(\omega)$) for $T=T_{0}=3$ K and the selected values of Coulomb pseudopotenital. We note, that the NDOS$(\omega)$ function is related to the $\Delta\left(T\right)$ solutions as:
\begin{equation}
\label{eq4}
{\rm NDOS}\left(\omega \right)=\frac{\rm DOS_{S}\left(\omega \right)}{\rm DOS_{N}\left(\omega \right)}={\rm Re}\left[\frac{\left|\omega -i\Gamma \right|}{\sqrt{\left(\omega -i\Gamma\right)^{2}}-\Delta^{2}\left(\omega\right)}\right],
\end{equation}
where $\rm DOS_{S}\left(\omega \right)$ and $\rm DOS_{N}\left(\omega \right)$ are respectively the density of states of the superconducting and normal state, whereas $\Gamma$ states for the pair breaking parameter and equals to $0.15$ meV.

As already mentioned, results presented in Fig. \ref{fig5} are depicted for $T=T_{0}$ due to the fact that only the maximum value of order parameter is relevant for our discussion. By comparing plots for different considered values of $\mu^{*}$, we can clearly observe the decrease of the superconducting gap towards metallic character of the discussed alloy. This outcome is in qualitative agreement with the results presented previously in Figs. \ref{fig1} (A-D).

On the basis of the obtained results we first estimate the value of the the zero temperature energy gap at the Fermi level ($2\Delta(0)\equiv2\Delta(T_{0})$). In the case of the PtH alloy at the pressure of 100 GPa the calculated values are: $2\Delta(0) \in \left< 6.87, 5.31, 4.28 \right>$ meV for $\mu^{*} \in \left< 0.05,  0.1,  0.15 \right>$. In the context of the BCS theory, the $2\Delta(0)$ parameter can be next used to determine the thermodynamic ratio for the zero temperature energy gap at the Fermi level ($R_{\Delta}$). In particular:

\begin{equation}
\label{r14}
R_{\Delta}\equiv\frac{2\Delta(0)}{k_{B} T_{C}}.
\end{equation}

Again, the BCS theory predicts constant value of the $R_{\Delta}$ ratio equals to 3.53, whereas our results for the PtH material under discussed conditions give $R_{\Delta} \in \left<  3.98, 3.84 \right>$ for $\mu^{*} \in \left< 0.05, 0.15 \right>$. In what follows, similarly like in the case of the $R_{H}$ and $R_{C}$ ratios, the $R_{\Delta}$ differ from the BCS suggestions.

%%%%%%%%%%%%%%%%%%%%%%%%%%%%%%%%%%%%%%%%%%%%%%%%%%%
\section{Summary}
%%%%%%%%%%%%%%%%%%%%%%%%%%%%%%%%%%%%%%%%%%%%%%%%%%%

In the present paper we address the vague experimental data for the high-pressure silane superconductor given by Erements {\it et al.} in \cite{eremets}, and the consequences of these results in the terms of the platinum hydride alloy. In particular, we perform calculations of the thermodynamic properties of the PtH hydride and reveal several crucial facts concerning possibility of the existence and properties of the superconducting phase in this material. In order to assure the creditability of our resutls, calculations are preformed at the pressure of 100 GPa and within the $hcp$ structure (with the P$6_{3}$ space group), which match exactly the conditions of the experiment conducted in \cite{eremets}.

In the first place, the calculated values of the critical temperature for three different $\mu^{*}$ present good agreement with the experimental data. In particular, it is shown that the experimental value of $T_{C}$ can be obtained for the $\mu^{*}$ equal to $\sim 0.1$. This is physically relevant value, which additionally agrees with the predictions of Ashcroft given for the hydrogen-dense materials in \cite{ashcroft1}. In what follows, our results indicate that experimental data obtained in \cite{eremets}, with a high probability, can be ascribed to the PtH alloy rather then SiH$_4$ material. This outcome is in agreement with the previous qualitative results for $T_C$ presented in \cite{errea2}, \cite{zhang}, \cite{zhou}, \cite{kim}. 

Our further calculations, allows us to determine other thermodynamic properties of the PtH at 100 GPa such as the thermodynamic critical field, the specific heat for the superconducting and normal state, the value of the energy gap at the Fermi level, and others.

It is convenient to summarize these results by estimating the values of the characteristic thermodynamic ratios. We show that all calculated ratios exceed predictions given by the BCS theory. In particular R$_{\rm H} \in \left< 0.155,  0.167 \right>$, $R_{C} \in \left< 1.88, 1.82 \right>$, and $R_{\Delta} \in \left<  3.98, 3.84 \right>$, which correspond to the range of $\mu^{*} \in\left< 0.05,  0.15 \right>$. The differences between our estimations and the Bardeen-Cooper-Schrieffer theory indicate a significant role of the retardation and strong-coupling effects in PtH at the pressure of 100 GPa. In what follows, the analysis of the discussed material cannot be conducted within the BCS theory, reinforcing the relevancy of our results and chosen theoretical approach. We note also, that the increase of the value of the Coulomb pseudopotential reduces mentioned discrepancies, which is the physically plausible effect.

%%%%%%%%%%%%%%%%%%%%%%%%%%%%%%%%%%%%%%%%%%%%%%%%%%%
\begin{acknowledgments}
Authors would like to thank R. Szcz{\c{e}}{\'s}niak (Cz{\c e}stochowa University of Technology, Poland) for the fruitful discussions and help throughout the work on the present article.
\end{acknowledgments}
%%%%%%%%%%%%%%%%%%%%%%%%%%%%%%%%%%%%%%%%%%%%%%%%%%%%%
\bibliographystyle{apsrev}
\bibliography{manuscript}
%%%%%%%%%%%%%%%%%%%%%%%%%%%%%%%%%%%%%%%%%%%%%%%%%%%%%
\end{document}